\documentstyle[11pt,fleqn]{article}
\newcommand{\nn}{\nonumber}            

\begin{document}
\parindent0mm

\newcommand{\reals}{\mbox{${\rm I\!R }$}}
\newcommand{\nats}{\mbox{${\rm I\!N }$}}
\newcommand{\intgs}{\mbox{${\rm Z\!\!Z }$}}
\newcommand{\komplex}{\mbox{${\rm I\!\!\!C }$}}
\newcommand{\beq}{\begin{eqnarray}}
\newcommand{\eeq}{\end{eqnarray}}
\newcommand{\sumh}{\sum_{\{P\}_{\Gamma}}}
\newcommand{\sume}{\sum_{\{Q\}_{\Gamma}}}
\newcommand{\rd}{(r^2+\delta^2)}
\newcommand{\erd}{\left(1+\frac{r^2}{\delta^2}\right)}
\newcommand{\sik}{\sin\left(\frac{k\pi} n\right)}
\newcommand{\iou}{\int_0^{\infty}}
\newcommand{\iuu}{\int_{-\infty}^{\infty}}
\newcommand{\siqk}{\sin^2\left(\frac{k\pi} n\right)}
\newcommand{\ch}{\chi^k(P)}
\newcommand{\chq}{\chi^k(Q)}
\newcommand{\sumkeu}{\sum_{k=1}^{\infty}}
\newcommand{\sumkem}{\sum_{k=1}^{n-1}}
\newcommand{\abl}{\partial}
\newcommand{\lm}{\ln\left(\frac{\lambda M^2}{2\mu ^2}\right)}
\newcommand{\slnuun}{\sum_{l_1,...,l_N=-\infty}^{\infty}
                 \!\!\!\!\!\!\!\!^{\prime}}
\newcommand{\klaus}{\sum_{l_1,...,l_{p-1}=-\infty}^{\infty}
                \!\!\!\!\!\!\!\!\!\!\!^{\prime}}



\title{Are p-Branes Asymptotically Black Holes ?}
\author{Andrei Bytsenko\thanks{Permanent address:
Department of Theoretical Physics,
State Technical University, St.~Petersburg 195251, Russia}\\
{\it Dipartimento di Fisica, Universit\`a degli Studi di Trento}\\
{\it 38050 Povo (Trento), Italia}\\
$$\\
Klaus Kirsten\thanks{E-mail: klaus@ebubecm1.bitnet. Alexander von
Humboldt Foundation Fellow.}\\
{\it Department E.C.M., Faculty of Physics, University of Barcelona}\\
{\it Diagonal 647, 08028 Barcelona, Spain}\\
$$\\
Sergio Zerbini\thanks{E-mail: zerbini@itncisca.bitnet}\\
{\it Dipartimento di Fisica, Universit\`a di Trento, Italia}\\
{\it Istituto Nazionale di Fisica Nucleare, Gruppo Collegato di Trento}}
\date{December 1993}
\maketitle
\thispagestyle{empty}
\begin{tabbing}
Subject classification number:\quad\= 1117\\
                                     \>0400
\end{tabbing}
\begin{abstract}
An attempt is made to compare the asymptotic state density of twisted
$p$-branes and the related state density of mass level $M$ of a
$D$-dimensional neutral black holes. To this aim, the explicit form
of the twisted $p$-brane total level degeneracy is calculated. The
prefactor of the
degeneracy, in contrast to the leading behaviour,
is found to depend on the winding number of the $p$-brane.
\end{abstract}
\vspace{3cm}
UB-ECM-PF 93/25\\
December 1993

\newpage
\setcounter{page}{1}
Recently, spacetimes with black $q$-brane solutions, namely singular
spacetimes for which the region of singularity assumes the shape of a
$q$-brane, have been constructed \cite{horowitzstrominger91}. Such
solutions have attracted much attention in view of the fact that they
can represent vacuum solutions of the 10-dimensional superstrings for
which $q=10-D$ is the dimensionality of an embedded flat space. The
problem of finding black $q$-brane solutions of the 10-dimensional
superstring theory can be reduced to the problem of finding black hole
solutions to the Einstein equations in $D$ dimensions
\cite{horowitzstrominger91,gibbonsmaeda88}. These solutions have been
used to study the statistical mechanics of black holes using the
microcanical ensemble prescription, this prescription being the unique
reasonable framework for analyzing the problem
\cite{harmsleblanc92,harmsleblancunv1,harmsleblancunv2}. For this
analysis, in refs.~\cite{harmsleblancunv1,harmsleblancunv2} the
approximate semiclassical formula for the neutral black hole degeneracy
$\rho(M)$ of states at mass level $M$ have been obtained, which
reads ($\hbar=c=1$, $G$ the generalized Newton constant)
\begin{eqnarray}
\rho (M) \sim B(M)\exp\left(S_E(M)\right)\label{1}
\end{eqnarray}
Here, $S_E$ is the Euclidean action (the so called Bekenstein-Hawking
entropy),
\begin{eqnarray}
S_E(M) =\sqrt{\pi}  \left[G \frac{2^{2D-3}}{(D-2)^{D-2}}
\Gamma\left(\frac{D-1} {2} \right)\right]
^{\frac{1}{D-3}}M^{\frac{D-2}{D-3}}\label{2}
\end{eqnarray}
and the prefactor $B(M)$ represents general quantum field theoretical
corrections to the state density unknown up to now. It is easily seen
that for $D=4$ the result reduces to the result found by Bekenstein
and Hawking
\cite{hawking75,hawking76}
\begin{eqnarray}
S_E(M) =4\pi G M^2\label{3}
\end{eqnarray}
The statistical mechanical density of states (degeneracies) given in
equation (\ref{1})-(\ref{3}) reveals great
similarities with the density $\rho_p (M)$ of states of higher
dimensional structures such as quantum $p$-branes, presented already
two decades ago in the form (assuming linear Regge-like
trajectories)
\cite{fubinihansonjackiw73,dethlefsennielsentze74,strumiaventuri75}
\begin{eqnarray}
\rho _p (M) \sim A(M)\exp\left(bM^{\frac{2p}{p+1}}\right)\label{4}
\end{eqnarray}
(see also \cite{alvarezortin92}), with some unknown factors
$A(M)$ and $b$, $b$ not depending on $M$.
More explicitly, comparison of
equation (\ref{2}) and (\ref{4}) yields
\cite{harmsleblancunv1}
\begin{eqnarray}
p=\frac{D-2}{D-4}\label{5}
\end{eqnarray}
The only solutions of equation (\ref{5}) with integer $p$ and integer
$D$ are given by $p=1$ ($D=\infty$),
that is the string case corresponds to an infinite dimensional black
hole,
$p=2$ ($D=6$), $p=3$ ($D=5$),
and
the limit $p\to\infty$ corresponds to the four dimensional black
hole \cite{alvarezortin92}.

Up to now, the comparison between the degeneracy of black holes and
$p$-branes is somehow vague, because, as shortly described, it is only
based on some part of the degeneracy, the rest of it not being
explicitly known. However, with the
help of the theorem of Meinardus \cite{meinardus54a,meinardus54b}, we
have calculated the complete $p$-brane state density, equation (\ref{4}),
including the up to now unknown factors $A(M)$ and $b$
\cite{bytsenkokirstenzerbini93}.
As we have shown there, $A(M)$ and
$b$ depend explicitly on $p$ and on the dimension $m$ of the embedding
space.
In that article, we
restricted however to the case of an equilateral torus with
compactification length $L=2\pi$ and
furthermore we did not take into
account the effects that winding numbers possibly have on the state
density.

Here, in order to make a further step toward the possible
identification of higher dimensional black holes as quantum excitation
mode of related p-branes  \cite{harmsleblancunv1}, we will generalize this
calculation to a $p$-brane with
arbitrary compactification lengths $L_i$ and winding numbers $b_i$,
with
$b_i=0$
or $1/2$. As we will proof, the leading exponential behaviour encoded in
the parameter $b$ don't depends on the winding numbers $b_i$. In
contrast the prefactor $A(M)$ strongly depends on the windings.

To begin with, we first calculate the complete asymptotic form
state density of a $p$-brane compactified on a manifold with the
topology $(S^1)^p\times R ^{m-p}$. We recall that the semiclassical
quantization of a $p$-brane in $(S^1)^p\times R ^{m-p}$, leads to the
"number operators" $N_{\vec n}$ with
$\vec n =(n_1,...,n_p)\in Z ^p$. The operators $N_{\vec n}$ and the
commutation relations for the oscillators can be found for example in
[15-17].
\nocite{bergshoeffsezgintownsend87,bergshoeffsezgintownsend87a}
\nocite{duffinamipopesezginstelle88}
The total number operator can be written as
\begin{equation}
N= \sum_{i=1}^d \sum_{\vec n \in Z^p/\{ 0 \}} w_{\vec n}N ^i_{\vec n}
\label{6}
\end{equation}
where $d=m-p-1$, the frequencies are given by
\begin{eqnarray}
w_{\vec n}^2 =\sum _{i=1}^p \left(\frac{2\pi\alpha^{\frac 1 {1+p}}
(n_i-b_i)}{L_i}\right)^2\label{7}
\end{eqnarray}
with the compactification lenghts $L_i$, $i=1,...,p$ of
the torus, the winding numbers $b_i=0$ or $1/2$,
and $\alpha$ is the $p$-brane tension with dimension {\it
mass}$^{-p-1}$.
For the purpose of generality let us deal with the number operator
\begin{eqnarray}
N=\sum_{i=1}^{d}\sum_{j }\omega_{j}N_{j}^i\label{general}
\end{eqnarray}
with $\omega_j^2=\lambda_j$, $\lambda_j$ being the eigenvalues of a
second order elliptic differential operator $L$ who's leading symbol is
the metric of a Riemannian manifold $M_p$ without boundary
(multiplied by $\alpha^{\frac 2 {1+p}}$ to make $\lambda_j$
adimensional).
In order to
evaluate the state density for the $p$-brane, one has to deal with the
trace of the heat number operator $\exp\{-tN\}$, the trace being computed
over the entire Fock space, namely
\begin{eqnarray}
Z(t) = {\mbox tr} e^{-tN}=\prod_{j}\left[1-e^{-t\omega_{j} }
\right]^{-d}\label{8}
\end{eqnarray}
where  $t>0$. For $p=1$, the function
$Z(z)$ of the complex variable $z=t+ix $ is known as the generating
function
of the partition function, which is well studied in the mathematical
literature \cite{hardyramanujan18}. These results have been used to
evaluate the asymptotic state density behaviour for $p=1$ [23-27].
\nocite{huangweinberg70,greenschwarzwitten87,mitchellturok87}
\nocite{mitchellturok87a,matsuo87}

In order
to state the result for arbitrary $p\in\nats$ let us introduce some
notations.
First we define
\begin{eqnarray}
\zeta_L(s)=\sum_j \lambda_j^{-s}\label{zeta}
\end{eqnarray}
to be the zeta function associated with
the  Laplace operator $L$, acting on the manifold $M_p$. For the torus
$M_p=T^p=(S^1)^p$,
the relevant zeta function is the Epstein zeta function
\cite{epstein03,epstein07}
\begin{eqnarray}
\zeta_L(\frac{s}{2})\equiv E_p(s;a_1,...,a_p;b_1,...,b_p)=\sum_{\vec n
\in Z ^p}\left[a_1(n_1-b_1)^2+...+a_p(n_p-b_p)^2\right]^{-\frac s
2}\label{9} \end{eqnarray}
where $a_i =\left(\frac{2\pi\alpha^{\frac 1
{1+p}}}
{L_i}\right)^2$,
and
the sum is defined for $\,\mbox{Re}\, s>p$. Here we assume that one of
the windings $b_i$ is not vanishing, otherwise the zero mode has
to be excluded from the sum in (\ref{9}).

It is well known that the zeta function $\zeta_L(\frac{s}{2})$ can
be analytically continued to a meromorphic function on the whole
complex plane with simple poles of order one at $s=p-r$,
$r=0,1,2,...$ with residues
$R_s=2C_{(p-s)/2}/((4\pi)^{\frac p 2}\Gamma(s/2))$.
Furthermore one has
$\zeta_L (-r)=(-1)^rr!C_{\frac p 2 +r}/(4\pi)^{\frac r 2}$.
Here $C_j$ are the Seeley-De
Witt coefficients defined as usual by
\cite{minakshisundarampleijel48,schwinger51,seeley67,seeley69,dewitt75}
\begin{eqnarray}
K(t)=\sum_j\,e^{-\lambda_jt}\sim \left(\frac 1 {4\pi t}\right)^{\frac
p 2}
\sum_{j=0}^{\infty}C_jt^j\label{11a}
\end{eqnarray}
especially one has $C_0=\alpha^{-p/(p+1)}Vol(M_p)$.
Furthermore, for $M_p=T^p$, we have
$E(0;a_1,...,a_p;$ $b_1,...,b_p
)=-1$ and $R_p =2\pi^{\frac p 2}/(\sqrt{\Delta}\Gamma(p/2))$, where
$\Delta =\prod_{i=1}^p a_i$. These properties do not depend
on the winding numbers $b_i$. As we will see, this leads to the
conclusion that the leading behaviour of the state density is
independent of the winding numbers.

Using these notations, a generalization of the theorem of Meinardus
leads to the following asymptotic expansion of the function $Z(z)$
for $z\to 0$ (the corresponding result
for  $M_p=T^p$ and $L_i=L$, $i=1,...,p$ in
\cite{bytsenkokirstenzerbini93} contains a typographical error)
\begin{eqnarray}
Z(z) \sim \exp\left\{d\left[\frac{B}
{z^p} -
\zeta_L(0)\ln z +\frac{1}{2}\zeta_L'(0)\right]\right\}\label{12}
\end{eqnarray}
with
\begin{eqnarray}
B=R_p\Gamma(p)\zeta_R(p+1)=\frac{Vol(M_p)\Gamma\left(\frac{p+1}
2\right)\zeta_R(p+1)}{\alpha^{\frac p {p+1}}
\pi^{\frac{p+1} 2}}\label{14a}
\end{eqnarray}
not depending on the winding numbers $b_i$.

In terms of $Z$, the total number $q(n)$ of $p$-brane states is
described by
\begin{eqnarray}
Z(z )=\sum_{n=0}^{\infty}q(n)e^{-zn}\label{13}
\end{eqnarray}
Adapting to our case the "thermodynamical methods" of ref.
\cite{fubinihansonjackiw73},
it is easy to arrive at the leading term of the
asymptotic expansion of $q(n)$ for $n\to\infty$. Infact $Z(t)$ may be
considered as a "partition function" and $t$ as the inverse
"temperature". The related "free energy" $F_t$, "entropy" $S_t$ and
"internal energy" $N_t$ may be written respectively as
\begin{eqnarray}
\qquad\qquad F_t=-\frac{1}{t}\ln Z(t)
\label{free}
\end{eqnarray}
\begin{eqnarray}
\qquad\qquad S_t=t^2\frac{\partial}{\partial t}F_t
\label{entropy}
\end{eqnarray}
\begin{eqnarray}
\qquad\qquad N_t=-\frac{\partial}{\partial t}\ln Z(t)
\label{number}
\end{eqnarray}
The limit $n\to\infty$ corresponds to $t \rightarrow 0$.
Futhermore in this limit
the "entropy" may be identified with $\ln q(n)$, while the "internal
energy" is related to $n$. Using (\ref{12}) one then finds
\begin{eqnarray}
\qquad\qquad F_t \simeq -dB t^{-p-1}
\label{100a}
\end{eqnarray}
\begin{eqnarray}
\qquad\qquad S_t \simeq d(p+1)B t^{-p}
\label{100b}
\end{eqnarray}
\begin{eqnarray}
\qquad\qquad N_t=dpB t^{-p-1}
\label{100c}
\end{eqnarray}
Eliminating the quantity $t$ between the last two equations, one gets
\begin{eqnarray}
\qquad\qquad S_t \simeq \frac{p+1}{p}(dpB)^{\frac{1}{p+1}}
N_t^{\frac{p}{p+1}} \label{100d}
\end{eqnarray}
As a result
\begin{equation}
\qquad\qquad \ln q(n) \simeq \frac{p+1}{p}(dpB)^{\frac{1}{p+1}}
n^{\frac{p}{p+1}} \label{100e}
\end{equation}
A more complete evaluation based on the result of Meinardus gives
\begin{eqnarray}
q(n) \sim C n^X\exp\left\{n^{\frac p {p+1}}\left(1+\frac 1
p\right)(dpB)^{\frac{1}
{p+1}}\right\}\label{14}
\end{eqnarray}
with the definitions
\begin{eqnarray}
C = e^{\frac{d}{2}\zeta_L'(0)}\left(2\pi (1+p)\right)^{-\frac 1 2}
(dpB)^{\frac{1-2d\zeta_L(0)}{2(1+p)}}\label{15}
\end{eqnarray}
and
\begin{eqnarray}
X=\frac{2d\zeta_L(0)-p-2}{2(p+1)}\label{16}
\end{eqnarray}
in which the complete form of the prefactor appears.
Assuming linear Regge trajectories we have the mass formula
$n=\alpha^{\frac 2 {p+1}}M^2$, the only consistent relation between
$n$ and $M$ in the semiclassical approximation scheme [17] (see also
[10])
and this yields
\begin{eqnarray}
\rho_p (M) \sim 2 C \alpha^{\frac y {1+p}}M^{y}
\exp\left\{\alpha^{2p}M^{\frac{2p}{p+1}}(\frac{p+1}{p}) [dpB] ^{\frac 1
{p+1}}\right\}\label{17}
\end{eqnarray}
with
\begin{eqnarray}
y=\frac{2p-2m+1}{1+p}\nonumber
\end{eqnarray}
Here one realizes that the leading behaviour of the state density
don't depends on the winding numbers. They are only entering in the
parameter $C$, (\ref{15}), due to $\zeta_L'(0)$. Furthermore it is seen
that this leading behaviour is very sensitive to the shape of the
$p$-brane model.
Eq.~(\ref{17}) may provide the result for comparison with the equation
(\ref{1})
and (\ref{2}) describing the black hole degeneracy of states at mass
level $M$.

Let us now consider the prefactor of equation (\ref{17}).
As is seen in equation (\ref{15}) we need some
information about $E'_p(0;a_1,...,a_p;b_1,...,b_p)$. Let us first
restrict to $b_1=...=b_p=0$ (we then write
$E_p(s;a_1,...,a_p;0,...,0)=E_p(s;a_1,...,a_p)$).
Using the analytical continuation
given in \cite{ambjornwolfram83}
\begin{eqnarray}
\hspace{-1.0cm}
E_p(s;a_1,...,a_p)&=&\frac 2 {a_p^{\frac s 2}} \zeta_R (s)
+\sqrt{\frac{\pi}{a_p}}\frac{\Gamma\left(\frac{s-1}
2\right)}{\Gamma\left(\frac s 2 \right)}E_{p-1}\left(s-1
;a_1,...,a_{p-1}\right)\label{epstein}\\
& &
\hspace{-1.0cm}
+\frac{4\pi ^{\frac s 2}}{\Gamma\left(\frac s 2\right)}\sqrt{\frac 1
{a_p}}\sum_{n=1}^{\infty}\sum_{n_1,...,n_{p-1}=-\infty}^{\infty}
\!\!\!\!\!\!\!\!\!\!\!\!'\hspace{1.0cm}
\left[\frac n
{\sqrt{a_p}\sqrt{a_1n_1^2+...+a_{p-1}n_{p-1}^2}}\right]^{\frac{s-1}
2}\times\nonumber\\
& &\hspace{-1.0cm}\qquad\qquad K_{\frac{s-1} 2}\left(\frac{2\pi
n}{\sqrt{a_p}}\sqrt{a_1n_1^2+...+a_{p-1}n_{p-1}^2}\right)\nonumber
\end{eqnarray}
(the prime means omission of the summation index $n_1=...=n_{p-1}=0$),
$K_s(z)$ are the modified Bessel functions,
the relevant information for equation (\ref{15}) reads
\begin{eqnarray}
\hspace{-1.0cm}
E_p'(0;a_1,...,a_p)&=&\frac 1 2
\ln\left(\frac{a_p}{4\pi^2}\right)-\frac{\pi}{\sqrt{a_p}}
E_{p-1}(-1;a_1,..,a_{p-1})\label{ablepstein}\\
& &
\hspace{-1.0cm}
-\sum_{n_1,...,n_{p-1}=-\infty}^{\infty}
\!\!\!\!\!\!\!\!\!\!\!\!'\hspace{1.0cm}
\ln\left(1-\exp\left\{-\frac{2\pi}{\sqrt{a_p}}
\sqrt{a_1n_1^2+..+a_{p-1}n_{p-1}n_{p-1}^2}\right\}\right)
\nonumber
\end{eqnarray}

With regards to the winding sectors of the theory one has the following
analytical continuation of the relevant Epstein zeta function (we assume
$b_p\neq 0$)
\cite{kirsten}
\beq
\lefteqn{E_p(s;a_1,...,a_p;b_1,...,b_p)=\sqrt{\frac{\pi^{p-1}}{a_1...
a_{p-1}}}\frac{\Gamma\left(\frac{s-p+1} 2\right)}{\Gamma\left(\frac s 2
\right)}a_p^{\frac{p-1-s} 2}\times}\nn\\
& &\left\{\zeta_H (s+1-p;1+b_p)+\zeta_H
(s+1-p;1-b_p)+b_p^{p-1-s}\right\}\nn\\
& &+\frac 2 {\Gamma(\frac s 2 )}\sqrt{\frac{\pi^{p-1}}{a_1...a_{p-1}}}
\klaus \hspace{.7cm}\sum_{l_p=-\infty}^{\infty}\exp\{2\pi
i(l_1b_1+...+l_{p-1}b_{p-1})\}\nn\\
& &\hspace{1cm}\times\left\{\frac{|l_p-b_p|}{\pi}\sqrt{\frac{a_p}
{\frac{l_1^2}{a_1}+...+\frac{l_{p-1}^2}{a_{p-1}}}}\right\}^{\frac{p-1-s}
2}\label{luci}\\
& &\hspace{1cm}\times K_{\frac{p-1-s} 2}\left(2\pi \sqrt{a_p} |l_p-b_p|
\sqrt{\frac{l_1^2}{a_1}+...+\frac{l_{p-1}^2}{a_{p-1}}}\right)\nn
\eeq
The prefactor is determined by the derivative of equation (\ref{luci}).
Due to the poles in the Gamma-functions, $p$ even and odd have to be
considered separately.  For $p$ even the result reads
\beq
\lefteqn{E_p'(0;a_1,...,a_p;b_1,...,b_p)=\frac 1 2
\sqrt{\frac{\pi^{p-1}}{a_1... a_{p-1}}}\Gamma\left(-\frac{p-1} 2\right)
a_p^{\frac{p-1} 2}\times}\nn\\
& &\left\{\zeta_H (1-p;1+b_p)+\zeta_H
(1-p;1-b_p)+b_p^{p-1}\right\}\nn\\
& &+\sqrt{\frac{\pi^{p-1}}{a_1...a_{p-1}}}
\klaus \hspace{.7cm}\sum_{l_p=-\infty}^{\infty}\exp\{2\pi
i(l_1b_1+...+l_{p-1}b_{p-1})\}\nn\\
& &\hspace{1cm}\times\left\{\frac{|l_p-b_p|}{\pi}\sqrt{\frac{a_p}
{\frac{l_1^2}{a_1}+...+\frac{l_{p-1}^2}{a_{p-1}}}}\right\}^{\frac{p-1}
2}\label{guido}\\
& &\hspace{1cm}\times K_{\frac{p-1} 2}\left(2\pi \sqrt{a_p} |l_p-b_p|
\sqrt{\frac{l_1^2}{a_1}+...+\frac{l_{p-1}^2}{a_{p-1}}}\right)\nn
\eeq
whereas for $p$ odd one finds
\beq
\lefteqn{E_p'(0;a_1,...,a_p;b_1,...,b_p)=
\frac{(-1)^{\frac {p-1} 2}}{\left(\frac{p-1} 2\right)!}
\sqrt{\frac{\pi^{p-1}}{a_1... a_{p-1}}}
a_p^{\frac{p-1} 2}\times}\nn\\
& &\left[\left\{\zeta_H'(1-p;1+b_p)+\zeta_H'(1-p;1-b_p)-b_p^{p-1}
\ln b_p\right\}\right.\nn\\
& &\left.+\frac 1 2 \left\{\psi\left(\frac{p+1} 2\right)+\gamma -\ln
a_p\right\}\left\{\zeta_H(1-p;1+b_p)+\zeta_H(1-p;1-b_p)+b_p^{p-1}\right\}
\right]\nn\\
& &+\sqrt{\frac{\pi^{p-1}}{a_1...a_{p-1}}}
\klaus \hspace{.7cm}\sum_{l_p=-\infty}^{\infty}\exp\{2\pi
i(l_1b_1+...+l_{p-1}b_{p-1})\}\nn\\
& &\hspace{1cm}\times\left\{\frac{|l_p-b_p|}{\pi}\sqrt{\frac{a_p}
{\frac{l_1^2}{a_1}+...+\frac{l_{p-1}^2}{a_{p-1}}}}\right\}^{\frac{p-1}
2}\label{sergio}\\
& &\hspace{1cm}\times K_{\frac{p-1} 2}\left(2\pi \sqrt{a_p} |l_p-b_p|
\sqrt{\frac{l_1^2}{a_1}+...+\frac{l_{p-1}^2}{a_{p-1}}}\right)\nn
\eeq
Let us mention that this might be slightly rewritten by using
$\zeta_H(1-p;a)=-B_p(a)/p$ with the Bernoulli polynomials $B_p(a)$
\cite{grad}.

We conclude with some remarks.
As already mentioned, comparison of equation (\ref{17}) with equation
(\ref{1}) and (\ref{2}) yields directly equation (\ref{5})
\cite{harmsleblancunv1}. Furthermore to find complete agreement of the
exponential function, one should fix the volume of the $p$-brane by
\begin{eqnarray}
\sqrt{\Delta}=\frac{\pi^{\frac p 2 -1}d\zeta_R(p+1)}{2^{3p-2}}
\frac{\Gamma(p+1)}{\Gamma\left(\frac p
2\right)\left(\Gamma\left(\frac{3p-1}{2p-2}\right)\right)^{p-1}}
\frac{(p+1)^{p+1}p^{p-1}}{(p-1)^{2p}}G^{1-p}
\alpha^{2p(p+1)}\label{condition}
\end{eqnarray}
It could give a relation between the generalized Newton constant $G$, the
generalized Regge slope parameter $\alpha$ and the volume of the
p-brane at the Planck scale.

It is seen that even for fixed $\Delta$ this leaves undetermined
$E_p'(0;a_1,...,a_p)$ and thus also the prefactor $C$, equation
(\ref{15}), in the state density (\ref{17}). So there is even no
correspondence between the asymptotics of these two models,
but one has to specify furthermore the compactification lenghts $L_i$.
For example, the
situation changes if we considered an equilateral torus
compactification $L_i=L$, $i=1,...,p$. The motivation for doing this
is, that probably due to the high symmetry of this compactification
the vacuum energy of the $p$-brane takes a minimum value (for example
this result has been found for a scalar field living on a torus
\cite{ambjornwolfram83}).
In this case, the quantity $E_p'(0;a_1,...,a_p)$, equation
(\ref{ablepstein}) takes a unique
value,
which determines uniquely the prefactor $A(M)$ in equation (\ref{4}).

With regards the role of the winding numbers, we have determined the
complicated dependence of the prefactor of the
asymptotic state density of the $p$-brane.
It is quite natural to investigate the
p-brane theory in arbitrary consistent curved backgrounds.
Unfortunately the geometric structure which describes these
backgrounds is very complicated (this is true also in the string
case). However, in contrast to the $p=1$ case (Riemann surfaces),
there are no satisfactory methods which permit to evaluate the
determinants of the Laplace operators acting on arbitrary flat vector
bundles over the (p+1)-manifolds. Attemps about the  semiclassical
quantization of p-branes on curved backgrounds of the form $ Ad S_4
\times S^7$ and $S^p\times S^1$ have been presented in
refs.~\cite{bergshoeffsalamsezgintanii88,nicolaisezgintanii88}.
As a consequence, anologous considerations can also be done in these
cases.

Finally, with regard to the fact that black holes may be viewed as
quantum excitation mode of the related p-branes, it is clear that the
presented letter is on a speculative level and
a further step
must be performed, namely a quantum mechanical evaluation of the black
hole prefactor  $B(M)$.  After this, one may draw some conclusion about
the possible relevance of p-branes in the physics of black holes.

\section*{Acknowledgements}
We thank G.~Venturi for helpful discussions. A.A.~Bytsenko thanks
I.N.F.N., gruppo collegato di Trento for hospitality and financial
support. K.~Kirsten thanks the members of the Department ECM, Barcelona
University, for the kind hospitality. Furthermore K.~Kirsten
acknowledges financial support from the Alexander von Humboldt
Foundation (Germany) and from CIRIT (Generalitat de Catalunya).

\end{document}